\documentclass[mathleft
]{an}
\usepackage{graphicx}
\usepackage{times}
\begin{document}

\Yearpublication{2011}%
\Yearsubmission{2010}%

\title{On the origin of gaseous galaxy halos - Low-column density gas in the Milky Way halo}

\author{N. Ben Bekhti\inst{1}\fnmsep\thanks{Corresponding author:
  \email{nbekhti@astro.uni-bonn.de}\newline}
\and  B. Winkel\inst{1}
\and  P. Richter\inst{2}
\and  J. Kerp\inst{1}
\and  U. Klein\inst{1}
}
\titlerunning{Gaseous galaxy halos}
\authorrunning{N. Ben Bekhti et al.}
\institute{
Argelander-Institut f\"{u}r Astronomie (AIfA)
Universit\"at Bonn
Auf dem H\"{u}gel\,71,
53121 Bonn, Germany 
\and 
Institut f\"ur Physik und Astronomie, 
Universit\"at Potsdam, 
Karl-Liebknecht-Str.\,24/25, 14476 Golm, Germany
}



\abstract{%
Recent observations show that spiral galaxies are surrounded by extended gaseous halos as predicted by the hierarchical structure formation scenario. The origin and nature of extraplanar gas is often unclear since the halo is continuously fueled by different circulation processes as part of the on-going formation and evolution of galaxies (e.g., outflows, galaxy merging, and gas accretion from the intergalactic medium). We use the Milky Way as a laboratory to study neutral and mildly ionised gas located in the inner and outer halo. Using spectral line absorption and emission measurements in different wavelength regimes we obtain detailed information on the physical conditions and the distribution of the gas. Such studies are crucial for our understanding of the complex interplay between galaxies and their gaseous environment as part of the formation and evolution of galaxies. Our analysis suggests that the column-density distribution and physical properties of gas in the Milky Way halo are very similar to that around other disk galaxies at low and high redshifts.  }

\maketitle

\section{Introduction}

Over the last years great progress has been made in understanding the properties and distribution of extraplanar gas around galaxies. Different gas phases are interacting with each other and influence the evolution of the host galaxy. Therefore, gaseous halos are perfect laboratories to study the aspects of (spiral) galaxy evolution. Using the full spectral regime it is possible to observe the different gaseous phases which helps to obtain a complete view of the galaxy--halo interaction. 

Extraplanar gas has been observed in emission around various spiral galaxies in form of neutral atomic hydrogen (e.g., NGC\,253 using the 21-cm line, Boomsma et al., 2005), hot X-ray gas (e.g., NGC\,253, Heesen et al., 2009), and diffuse ionised gas (via H$_\alpha$, e.g., NGC\,5775, Rossa \& Dettmar, 2003). These observations suggest a multiphase medium in the halo of galaxies where warm diffuse and compact cold cloud-like objects are embedded in a hot gaseous galactic corona. 

X-ray telescopes are able to detect the hot $T \approx 10^6$\,K gas beyond the galactic disk (e.g., Pietz et al., 1998; Nicastro, Mathur \& Elvis, 2008). Measurements of \ion{O}{vi} absorption lines in the direction of QSOs confirmed the presence of a hot gaseous galactic halo (Sembach et al., 2003). 
Rossa \& Dettmar (2003) showed with their $H_\alpha$ survey that extraplanar diffuse ionised gas is present in all galaxies with a certain star formation rate.

Optical and ultraviolet absorption spectroscopy provides a sensitive tool to analyse different species constituting the halo gas (for a review, see Richter 2006). The observed ionisation stages range from low-ionisation (\ion{H}{i}, \ion{Na}{i}, \ion{Ca}{ii}, \ion{Mg}{ii}) to highly ionised gas traced by \ion{O}{vi} \ion{Si}{iv}, etc. Especially, \ion{Mg}{ii} absorption against quasars (QSOs) was extensively used to analyse the absorption characteristics of extragalactic halo structures (e.g., Charlton, Churchill, \& Rigby, 2000; Bouch\'{e} et al., 2006). Absorption spectroscopy has the advantage that it is very sensitive to low-column densities and that it is independent of the distance to the absorbing objects as long as the background continuum source is bright enough.

Spectacular examples for galaxies with extended gaseous \ion{H}{i} halos are the nearby spiral galaxies NGC\,891 and NGC\,262 (e.g., Swaters, Sancisi, \& van der Hulst, 1997; Oosterloo, Fraternali, \& Sancisi, 2007). NGC\,891 is one of the best-studied edge-on galaxies. Observations in different wavelength regimes (Whaley et al., 2009, and references therein) revealed the existence of an extended radio halo, an extended layer of diffuse ionised and hot-ionised gas (e.g., Bregman \& Pildis, 1994). Deep \ion{H}{i} observations with the Westerbork Radio Sythesis Telescope (WSRT) were performed by Oosterloo, Fraternali, \& Sancisi, 2007 and they found that 25\% of the total \ion{H}{i} mass resides in the halo. The neutral gas extends to large heights of $z \approx 8\ldots10$\,kpc. A long filament extends even up to $z\sim20\mathrm{\,kpc}$. In the case of NGC\,262, a  past interaction and gas accretion is probable as suggested by the presence of an enormous \ion{H}{i} envelope with 176\,kpc in diameter and a large tail-like extension. 
In the case of NGC\,891, NGC\,262 and many other spiral galaxies in the Local Volume, the observations indicate that up to 30\% of the total \ion{H}{i} mass is situated in the halo. The gas in these galaxies is lagging behind the rotation of the host galaxy by up to $20\,\mathrm{km\,s}^{-1}$ and shows a global infall motion.

Our own galaxy, the Milky Way, is surrounded by various gaseous structures as well. The most prominent gaseous objects in the Milky Way halo are the so-called intermediate- and high-velocity clouds (IVCs, HVCs). This population of gas clouds has typically \ion{H}{i} column densities of more than $N_\mathrm{HI} \approx 10^{19}$\,cm$^{-2}$ and radial velocities inconsistent with a model of differential rotation of the galactic disk. Besides the velocity there are other criteria for the distinction between IVCs and HVCs. IVCs are relatively nearby objects with typical distances to the disk of $z \leq 2$\,kpc and metallicities of 0.7 to 1.0  whereas HVCs are located at larger distances ($z \leq 50$\,kpc) and have metallicities of 0.1 to 1.0 (e.g., Wakker, 2001; Wakker, et al., 2008; Richter et al., 2001). 

IVCs and HVCs can be detected all over the sky but they are not homogeneously distributed. On one hand, there are extended coherent complexes (like complex A, C, and M) on the other hand one observes large streams (like the Magellanic stream) spanning tens of degrees on the sky. Furthermore, there are numerous isolated and compact clouds (so-called Compact High-Velocity Clouds, CHVCs). A problem of IVC/HVC research is the lack of accurate distance measurements which result in a poor constraint of physical parameters like mass, particle density, and size of the clouds. To set limits, stellar spectroscopy can be used in some cases to find distance brackets (e.g., Thom et al., 2008; Wakker et al. 2007, 2008; Richter et al. 2001) and for a few large features kinematic analyses could be successfully applied (e.g., in case of Smiths cloud; Lockman et al., 2008; or for the Magellanic System; Gould, 2000). 

After the discovery of the Milky Way HVC population the question emerged, whether IVCs and HVCs are a common phenomenon around spiral galaxies. Thilker et al. (2004) used the Green Bank Telescope (GBT) to search for this cloud population around M\,31. They found 20 discrete features located within 50\,kpc of the M\,31 disk with line widths in the range of $10\ldots70$\,km\,s$^{-1}$ and typical \ion{H}{i} column densities of $N_\mathrm{HI} \approx 10^{19}$\,cm$^{-2}$ showing the same physical properties as Milky Way IVCs and HVCs. Westmeier, Br\"{u}ns, \& Kerp (2008) mapped a large area around M\,31 in 21-cm line emission with the Effelsberg telescope. Their survey extends out to a projected distance of about 140\,kpc. The goal was to search for neutral gaseous structures beyond 50\,kpc. 
The nondetection down to an \ion{H}{i} column density detection limit of $N_\mathrm{HI}= 2.2 \cdot 10^{18}$\,cm$^{-2}$ (corresponding to $8\cdot10^4$ solar masses) suggests that IVCs and HVCs are generally found in the proximity of their host galaxies.
A key advantage of studying the gaseous halo objects around M\,31 is that their distance can be well-constrained which enables the estimation of distance-dependent parameters and the area filling factor of the gas. In case of M\,31 the area filling factor is about 30\% where the concentration of the \ion{H}{i} gas is decreasing with increasing radius (Richter et al., in prep.).

The properties of the extraplanar gas are manifold. There is a large variety of metal abundances, densities, and ionisation states. This observations make a single origin of the gas unlikely. Today, four major origin scenarios are favored which can be divided into galactic and extragalactic origin. The Galactic fountain model which was proposed by Shapiro \& Field (1976) explains the extraplanar gas as a result of supernovae explosions which cause outflows of metal-enriched gas from the disk into the halo. There it cools, condenses and falls back onto the disk. Alternatively, halo gas can be the result of stellar winds of massive stars (e.g., Martin, 2006). Oort, 1966 suggested that the gas is of primordial origin and represents the leftovers of the early galaxy formation. 

Finally, major and minor interaction processes between galaxies (and also between galaxies and their environment) expel gas into the halo via ram-pressure interaction or tidal stripping (Gardiner \& Noguchi, 1996). Such interactions are probably also the reason for the observed warping of the outer neutral hydrogen gas layers (e.g., Sancisi et al., 1976; Bottema, 1996) and the lopsidedness (e.g., Sancisi et al., 2008, and references therein) of disk galaxies. Obviously, all these effects substantially influence the evolution of the host galaxy. The complex interplay between the different processes make it often complicated from the observers point of view to relate the various structures seen in the halos of disk galaxies to a certain origin. 

The mass and energy exchange between the disk and halo is in many ways fundamental for the galactic life cycle. In order to reach the accretion rates to sustain the observed constant star formation rates for disk galaxies in the Local Volume (Brinchmann et al., 2004), additional infall of large amounts of fresh gas from the Intergalactic Medium (IGM) is necessary. 

Another important aspect is that the halo gas is believed to represent the interface between the condensed galactic disk (well observed) and the surrounding IGM (not well observed) (e.g., Fraternali et al., 2007). 
CDM cosmology predicts that most of the baryonic matter in the local universe is in the IGM (Cen \& Ostriker, 2006). Because of the complexity and the many physical processes taking place our knowledge of the IGM is still incomplete. Studying the exchange of material between galaxies and their environment is therefore an efficient way to probe the IGM.

\section{Motivation of our project}

In our project we use the Milky Way as a laboratory to systematically analyse the low-column density halo gas. 
Although there is a large amount of available 21-cm emission data from large \ion{H}{i} surveys like the Leiden-Agentine-Bonn survey (LAB; Kalberla et al., 2005), the Galactic All-Sky survey (GASS; McClure-Griffiths et al.,2009; Kalberla, P.M.W., et al., 2010), and the new Effelsberg-Bonn \ion{H}{i} survey (EBHIS; Kerp, 2009; Winkel et al., 2010a) they are all limited to column densities above $N_{\ion{H}{i}} \simeq 10^{18}$\,cm$^{-2}$. Additionally, the relatively poor angular resolution makes it hard to detect low-column density small-scale structures.
The solution to this issue is to use (metal) absorption line spectroscopy against QSOs, which is much more sensitive to low-column density gas (Richter et al., 2009).

Almost all recent absorption studies of IVCs and HVCs were carried out in the UV to study the metal abundances and ionisation conditions of halo clouds. These studies were designed as follow-up absorption observations of known IVCs and HVCs, thus representing an 21-cm emission-selected data set. 
To statistically compare the absorption characteristics of the extraplanar gas with the properties of intervening metal-absorption systems towards QSOs one requires an absorption-selected data set of IVCs and HVCs. 

Our analysis will help us to answer the question whether the known IVCs and HVCs around the Milky Way represent most of the neutral gas mass in the halo or are just the tip of the iceberg of what is observed in 21-cm emission today.

\section{Data}
Our absorption sample consists of 408 archival (Spectral Quasar Absorption Database, SQUAD; PI: M.~T. Murphy) high-resolution ($R \approx 40000\ldots60000$, corresponding to $6.6\,\mathrm{km\,s}^{-1}$\,FWHM) QSO spectra observed with VLT/UVES. We searched for \ion{Na}{i} and \ion{Ca}{ii} (\ion{Ca}{ii}$\lambda 3934.77, 3969.59\,\AA$ and \ion{Na}{i}$\lambda 5891.58,5897.56\,\AA$) absorption of Milky Way halo gas. \ion{Ca}{ii} and \ion{Na}{i} with their low ionisation potentials of 11.9\,eV and 5.1\,eV are trace species of cold and warm neutral gas.

The absorption measurements were complemented with 21-cm single-dish \ion{H}{i} emission line data using the new more sensitive \ion{H}{i} surveys EBHIS ($\Theta= 9\arcmin$ HPBW, velocity channel separation $\Delta v=1.3$\,km\,s$^{-1}$) and GASS ($\Theta= 14\arcmin$ HPBW, velocity channel separation $\Delta v=0.8$\,km\,s$^{-1}$) to search for neutral hydrogen connected with the absorption lines. Both surveys have a column density detection limit of about $N_\mathrm{HI} \approx 3\cdot10^{18}\,\mathrm{cm}^{-2}$ (calculated for a line width of $20\,\mathrm{km\,s}^{-1}$.) Furthermore, for several sight lines we performed pointed observations with integration times of 15\,min with the 100-m telescope Effelsberg. While the latter allow for the detection of lower column densities, the \ion{H}{i} survey data provide the opportunity to study the gaseous environment of the detected absorbers.

Having such a large data sample allows for the first time to systematically study the global properties and distribution of the neutral (low-column density) Milky Way halo gas. 

To decide whether the detected clouds participate in galactic rotation or not, we used a kinematic model of the Milky Way (Kalberla et al., 2003; Kalberla et al., 2007) and calculated the deviation velocities (Wakker, 1991) for each component. 
In 126 (75) lines of sight we detect 226 (96) \ion{Ca}{ii} (\ion{Na}{i}) absorption components at intermediate- and high-velo\-cities (Ben Bekhti et al., 2008).
Along 100 sight lines (EBHIS and Effelsberg: 38, GASS: 62) the \ion{Ca}{ii} and/or \ion{Na}{i} absorption is connected with \ion{H}{i} gas. 
\begin{figure*}
\centering
\includegraphics[height=0.58\textwidth, angle=90,bb=95 533 928 914, clip=]{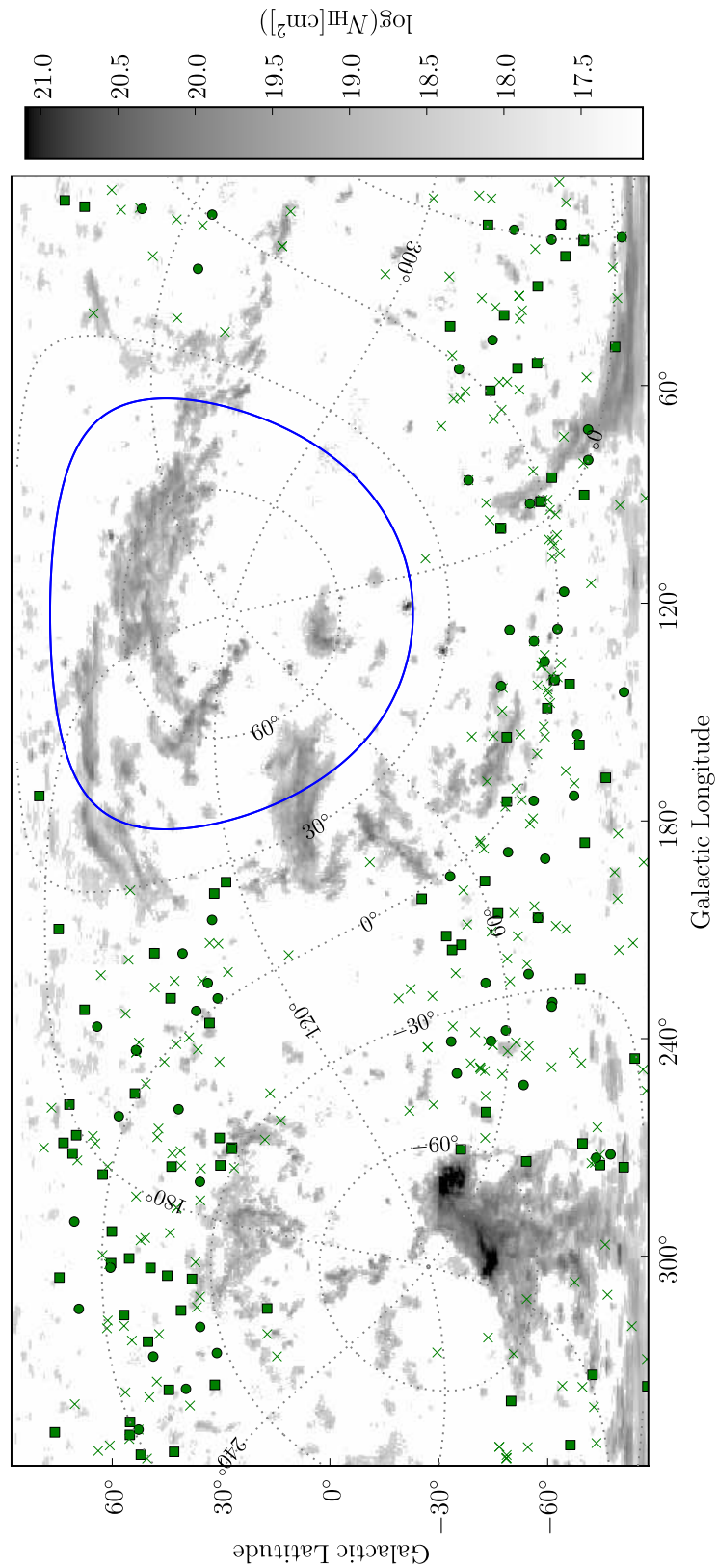}
\caption{HVC-all-sky map calculated by T.~Westmeier based on the data of the Leiden-Argentine-Bonn survey (LAB, Kalberla et al., 2005). The different symbols mark the positions of 408 sight lines that were observed with VLT/UVES. Along 126 (75) lines of sight we
detect \ion{Ca}{ii} (\ion{Na}{i}) absorption components (marked with circles and boxes). The boxes mark the positions where we found corresponding \ion{H}{i} emission with the 100-m telescope at Effelsberg (38~sight lines) or from the GASS data (62~sight lines). The crosses indicate non-detections.}
\label{HVC_allsky}
\end{figure*}

Figure\,\ref{HVC_allsky} shows an all-sky HVC map based on the data of the LAB survey.
The different symbols mark the positions of 408 sight lines that were observed with VLT/UVES. Detected absorption components are marked with circles. The boxes indicate sight lines were we find corresponding \ion{H}{i} emission lines with EBHIS and/or GASS. The crosses show non-detections.
Almost 50\% of the intermediate- and high-velocity \ion{Ca}{ii} and \ion{Na}{i} components might be associated to known IVC/HVC complexes considering their spatial position and radial velocity.

\begin{figure*}
\centering
\includegraphics[width=0.47\textwidth]{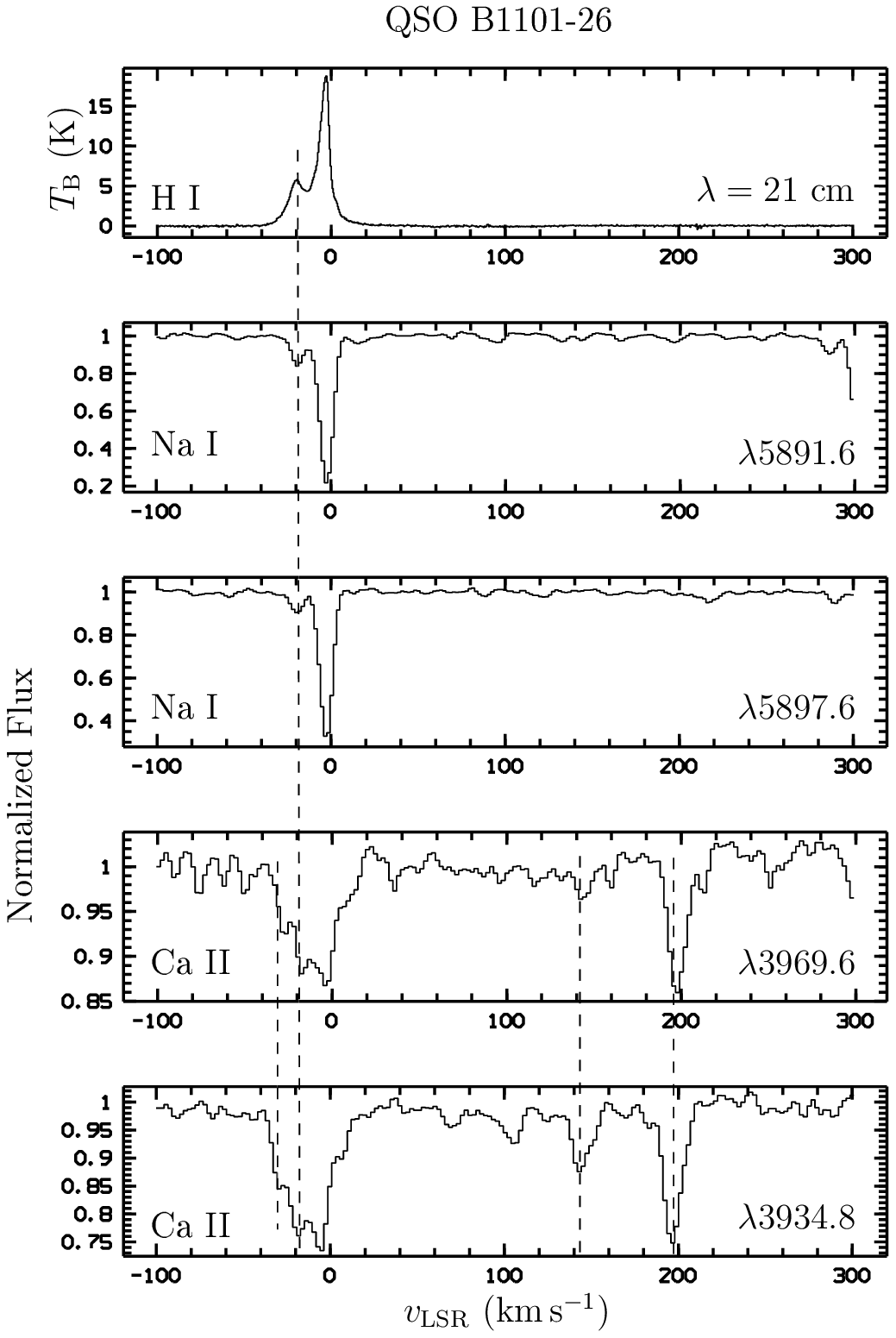}
\includegraphics[width=0.47\textwidth]{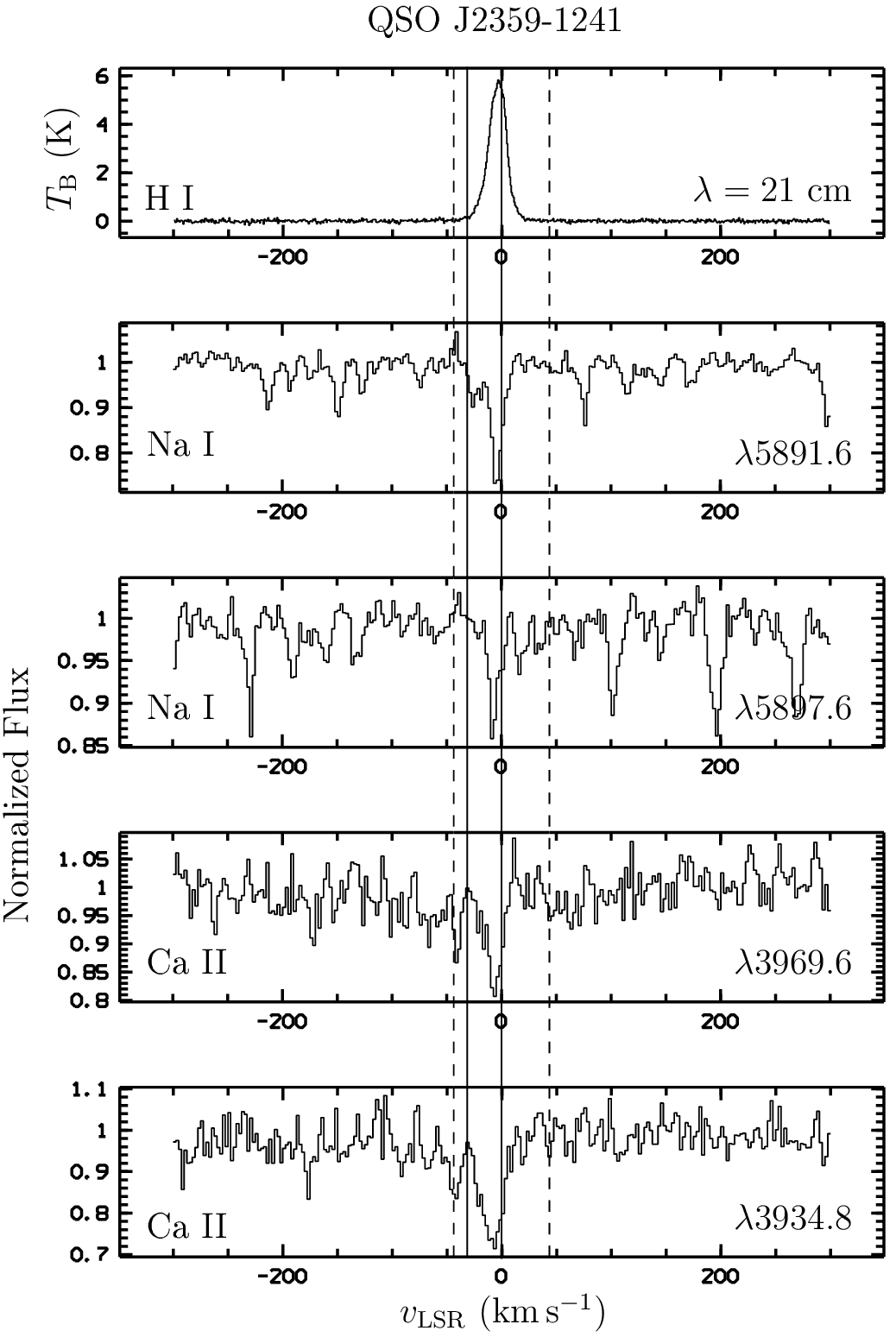}
\caption{Example spectra of the quasars QSO\,B1126$-$26 and QSO\,J2359$-$1241. The absorption and corresponding emission lines are indicated by dashed lines. The solid lines mark the $v_\mathrm{lsr}$ velocity range expected for the Milky Way gas in that direction according to a model by Kalberla et al. (2003,2007).}
\label{spectrum}
\end{figure*}
Figure\,\ref{spectrum} displays two example spectra with optical absorption of the \ion{Ca}{ii} and \ion{Na}{i} doublets. Additionally, the \ion{H}{i} 21-cm emission line spectra are shown. The solid lines mark the minimal and maximal radial velocities expected for the Galactic disk gas according to the Milky Way model.
The dashed lines indicate the location of the absorption and corresponding emission lines.
It is remarkable that in many cases one observes distinct absorption lines, but no corresponding 21-cm emission is seen. This suggests that either the \ion{H}{i} column densities are below the \ion{H}{i} detection limits or that the diameters of the absorbers are very small such that beam-smearing effects render them undetectable.

\section{Results}

We performed a detailed statistical analysis using the large data sample (Ben Bekhti et al., 2008; Ben Bekhti et al., in prep.). In the following we will present some of the results.

The absorption systems have \ion{Ca}{ii} and \ion{Na}{i} column densities in the range of $N_\ion{Ca}{ii} \approx 7.8 \cdot 10^{10}\mathrm{cm}^{-2} \ldots 1 \cdot 10^{14}\mathrm{cm}^{-2}$ and $N_\ion{Na}{i} \approx 3.2 \cdot 10^9\mathrm{cm}^{-2} \ldots 1.3 \cdot 10^{13}\mathrm{cm}^{-2}$ and Doppler parameters of $b < 7$\,km\,s$^{-1}$ (median value: $b \sim 3$\,km\,s$^{-1}$). From the Doppler parameters  an upper kinetic temperature limit of $T_\mathrm{max} \leq 1.2 \cdot 10^5$\,K ($T_\mathrm{max} \leq 2 \cdot 10^4$\,K) can be calculated showing that the line widths are likely enhanced due to turbulent effects, as at least \ion{Na}{i} usually traces the cold and dense cores of the clouds. From the EBHIS and GASS \ion{H}{i} data we get column densities in the range of 
$N_\ion{H}{i} \approx 1 \cdot 10^{19}\mathrm{cm}^{-2} \ldots 1 \cdot 10^{20}\mathrm{cm}^{-2}$ and Doppler parameters of $b < 20$\,km\,s$^{-1}$ leading to an upper kinetic temperature limit of $T_\mathrm{max} \leq 2 \cdot 10^4$\,K which is typical for warm neutral gas clouds observed in the halo.

\begin{figure*}
\centering
\includegraphics[width=0.47\textwidth]{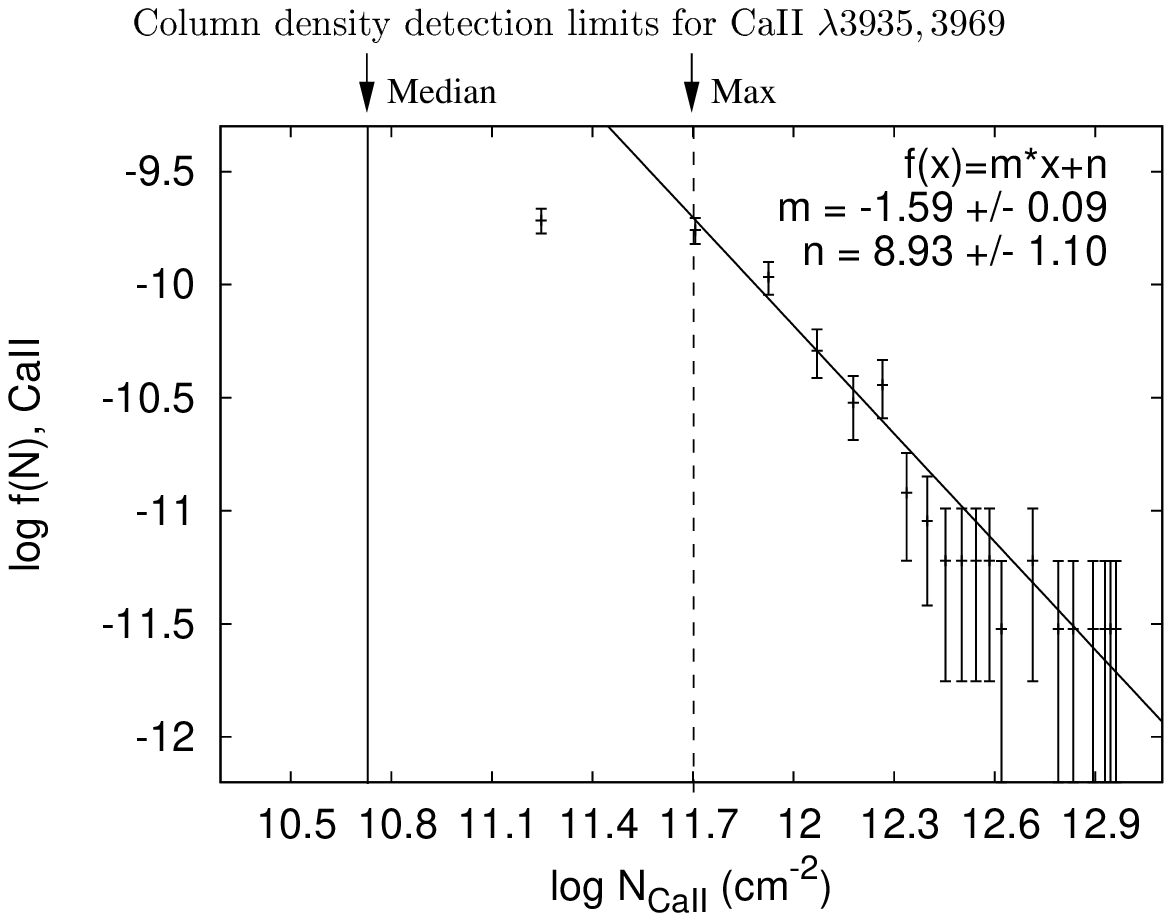}
\includegraphics[width=0.47\textwidth]{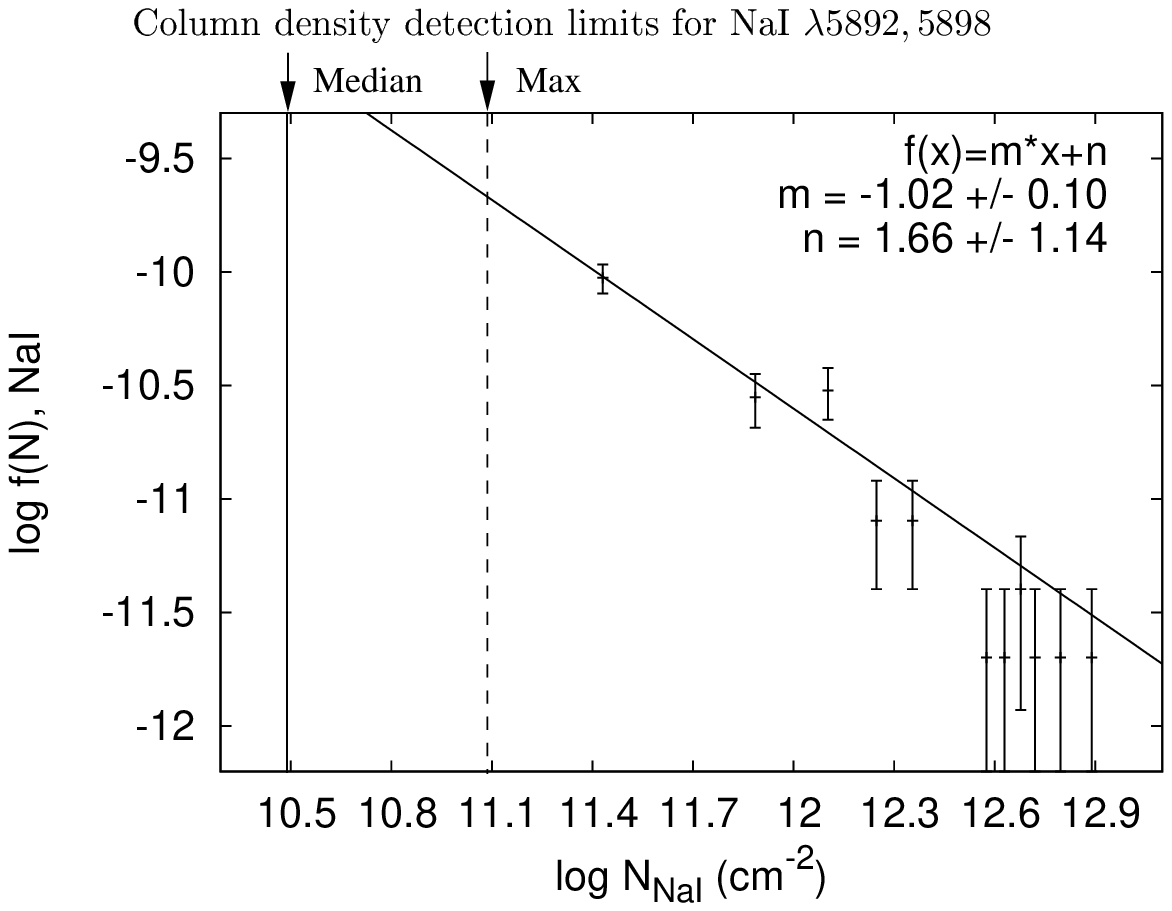}
\caption{The \ion{Ca}{ii} and \ion{Na}{i} column density distributions (CDD), $f(N)$, as derived from the VLT/UVES data, fitted with a power-law  $f(N)=CN^{\beta}$. We obtain $\beta_\ion{Ca}{ii}=-1.6 \pm 0.1$, $\log C_\ion{Ca}{ii}=8.9 \pm 1.1$ and  $\beta_\ion{Na}{i}=-1.0 \pm 0.1$, $\log C_\ion{Na}{i}=1.7 \pm 1.1$. The vertical solid lines indicate the $4 \sigma$ detection limit $\log (N_\ion{Ca}{ii}, N_\ion{Na}{i}[\mathrm{cm}^{-2}])=(10.7, 10.4)$ for the median $S/N_\mathrm{r}$ and the dotted lines represent the detection limit $\log (N_\ion{Ca}{ii}^\mathrm{max}, N_\ion{Na}{i}^\mathrm{max}[\mathrm{cm}^{-2}])=(11.7, 11.1)$ for the lowest $S/N_\mathrm{r}$ spectra. The latter leads to an incompleteness in the source catalog, hence, the power-law fits were applied to values $\log N_\ion{Ca}{ii} > 11.7$\,cm$^{-2}$ and $\log (N_\ion{Na}{i}/\mathrm{cm}^{-2})>11.1$, respectively.}
\label{CDD_caII}
\end{figure*}

Figure\,\ref{CDD_caII} shows the column density distribution (CDD) function (Churchill, Vogt \& Charlton, 2003) of the \ion{Ca}{ii} and \ion{Na}{i} absorbers derived from the VLT/UVES data. The \ion{Ca}{ii} (\ion{Na}{i}) CDD follows a power law $f(N)=CN^{\beta}$ with $\beta=-1.6 \pm 0.1$ ($\beta=-1.0 \pm 0.1$) for $\log N_\ion{Ca}{ii} > 11.6$\,cm$^{-2}$ ($\log (N_\ion{Na}{i}/\mathrm{cm}^{-2})>11.1$). The vertical solid lines indicate the median detection limit in our sample. The dotted lines represent the spectrum with the highest noise level according to the worst detection limit leading to incompleteness of the sample below the associated column density values. The flattening of the \ion{Ca}{ii} and \ion{Na}{i} distributions towards lower column densities can be attributed to this selection effect.

Churchill, Vogt, \& Charlton (2003) derived the CDD function for strong \ion{Mg}{ii} systems in the vicinity of other galaxies at redshifts $z=0.4 \ldots 1.2$ and found a slope of $\beta=-1.6 \pm 0.1$ which is in good agreement with our \ion{Ca}{ii} CDD. \ion{Ca}{ii} and \ion{Mg}{ii} have comparable chemical properties and  both trace neutral gas in halos. The fact that both slopes agree quite well suggests that the statistical properties of the halo absorption-line systems are similar at low and high redshifts.

\begin{figure}
\centering
\includegraphics[width=0.48\textwidth]{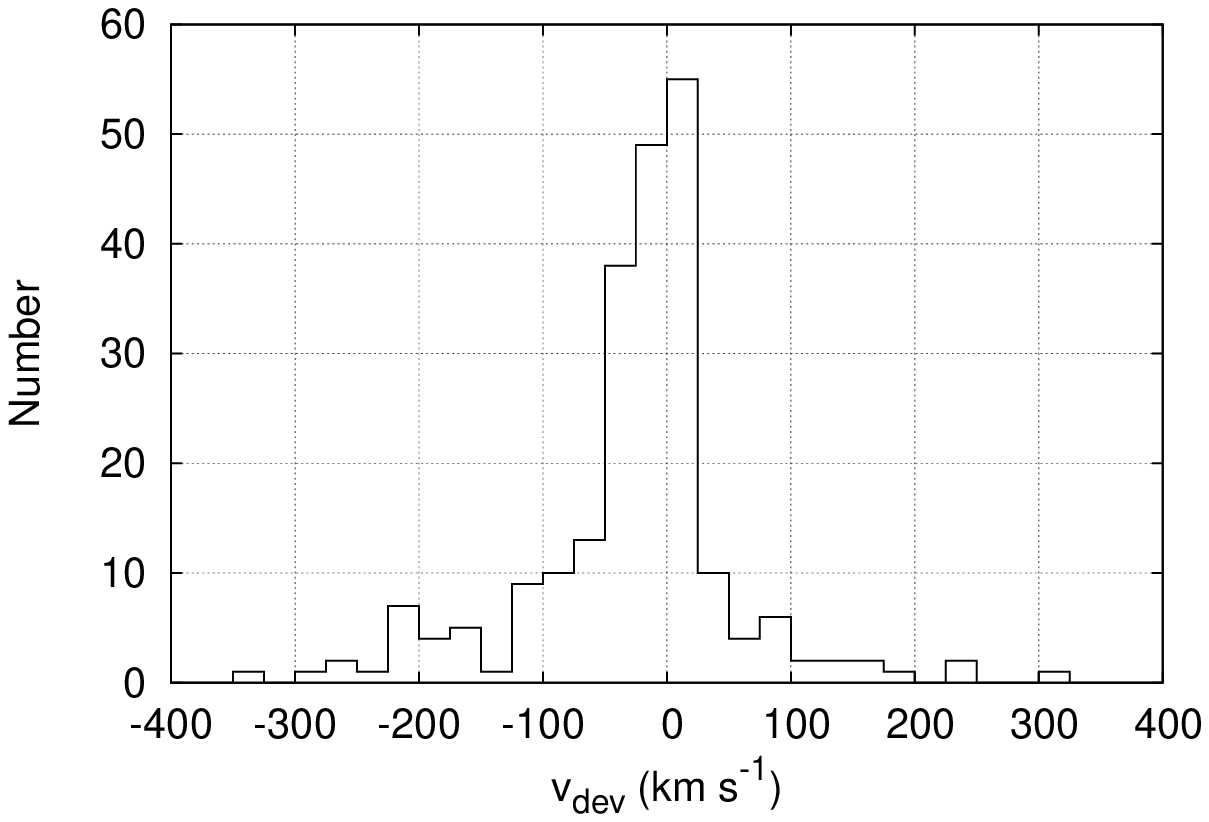}\\[1ex]
\includegraphics[width=0.48\textwidth]{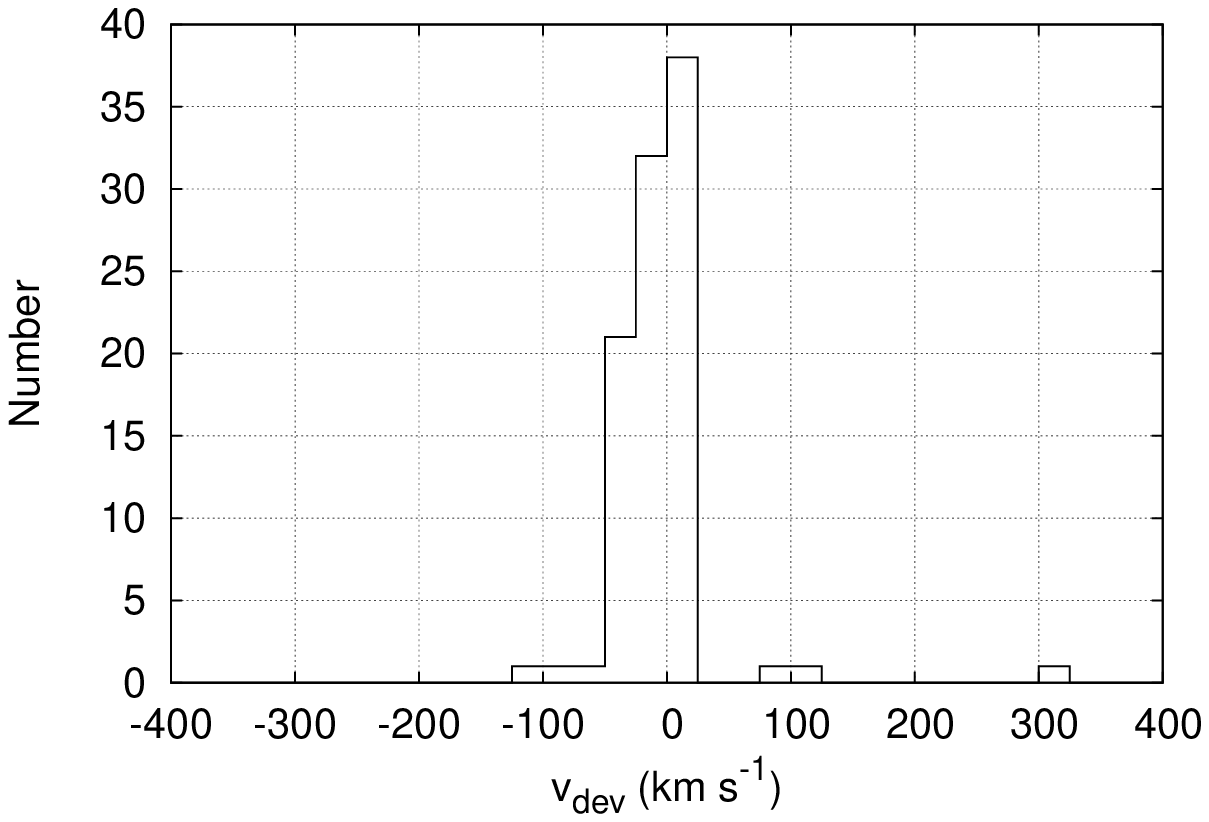}
\caption{Number of \ion{Ca}{ii} and \ion{Na}{i} intermediate- and high-velocity halo absorbers as a function of deviation velocity, $ v_\mathrm{dev}$.}
\label{vdev_histo_caII}
\end{figure}

Figure\,\ref{vdev_histo_caII} shows the number of \ion{Ca}{ii} and \ion{Na}{i} absorbers as a function of deviation velocity. The bulk of absorbing clouds has deviation velocities of $ \vert v_\mathrm{dev} \vert < 50$\,km\,s$^{-1}$ with a notable excess towards negative deviation velocities. This fits to previous studies of extraplanar gas indicating an excess of clouds infalling towards the Galactic disk (Oosterloo, Fraternali \& Sancisi, 2007, and references therein).

\begin{figure}
\centering
\includegraphics[width=0.48\textwidth,bb=0 0 495 381]{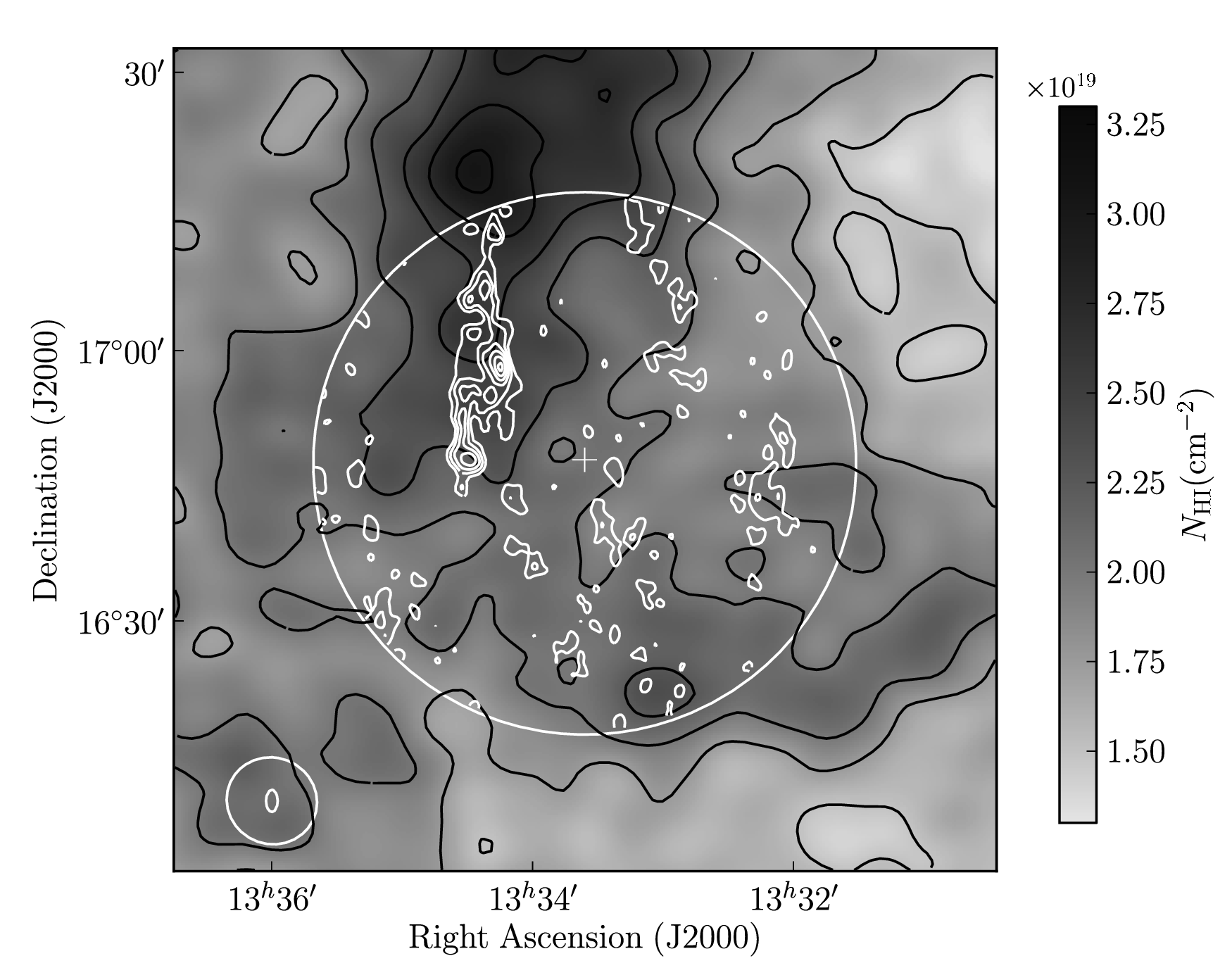}\\[2ex]
\includegraphics[width=0.48\textwidth,bb=0 0 495 381]{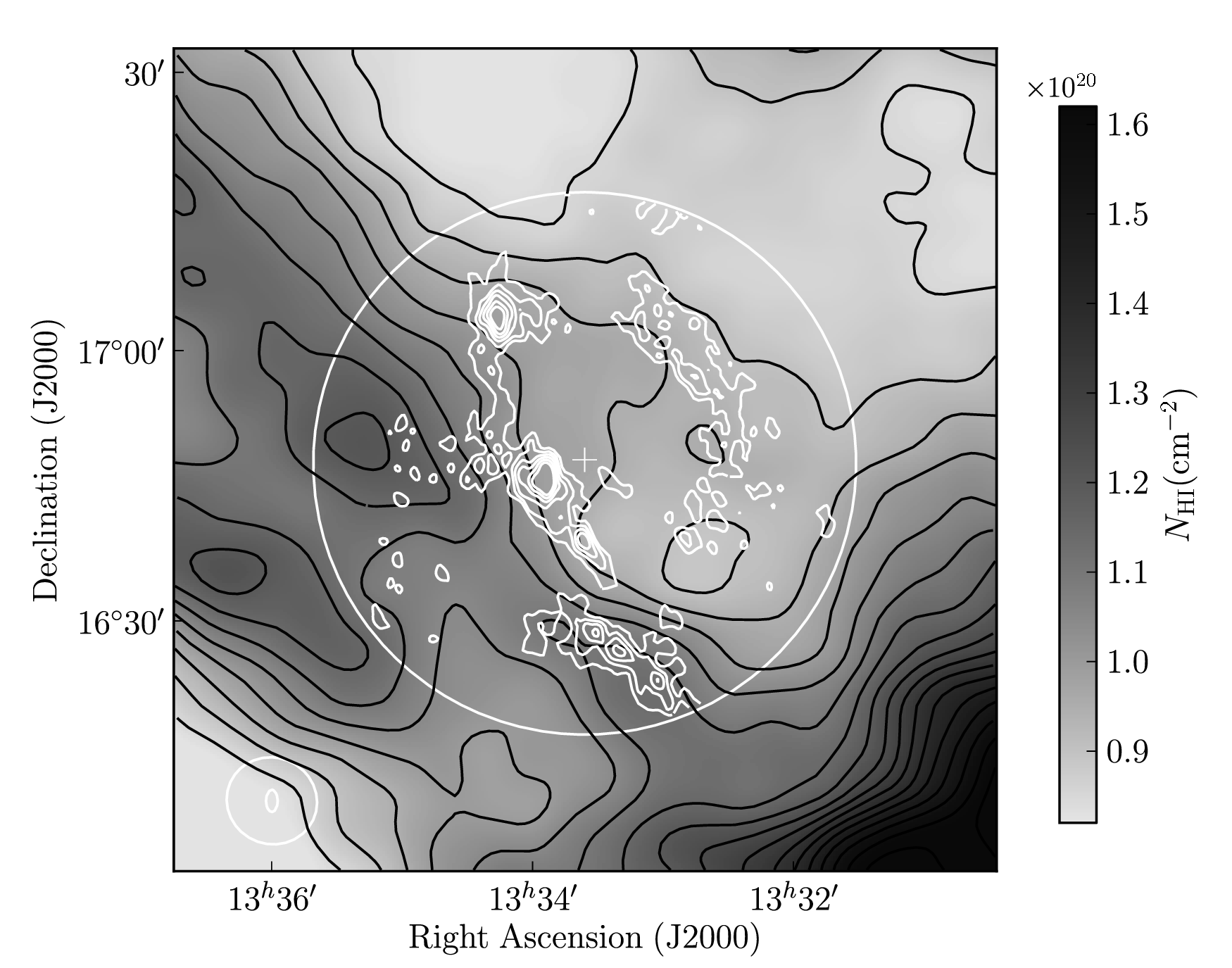}
\caption{\ion{H}{i} column density map of the two velocity components (top panel: $v_\mathrm{lsr}\sim-27\,\mathrm{km\,s}^{-1}$, bottom panel: $v_\mathrm{lsr}\sim-9\,\mathrm{km\,s}^{-1}$) in the direction of QSO\,B1331$+$170 as observed with EBHIS. The black contours start at $N_\ion{H}{i} = 1.5\cdot 10^{19}\,\mathrm{cm}^{-2}$ and increase in steps of $2.5\cdot 10^{18}\,\mathrm{cm}^{-2}$ (top panel) and $N_\ion{H}{i} = 8.5\cdot 10^{19}\,\mathrm{cm}^{-2}$ in steps of $5\cdot 10^{18}\,\mathrm{cm}^{-2}$ (bottom panel). Overlaid (white) are the contour lines of the high-resolution observation made with the WSRT, starting at $N_\ion{H}{i} = 1\cdot 10^{18}\,\mathrm{cm}^{-2}$ in steps of $1\cdot 10^{18}\,\mathrm{cm}^{-2}$. In the lower left corner the beam sizes of EBHIS and the WSRT (synthesized beam) are indicated. The line of sight towards QSO\,B1331$+$170 is marked with a white cross, the white circle denotes the size of the primary beam of the WSRT ($2\times\mathrm{HPBW}$).}
\label{ebhis_WSRT}
\end{figure}

\begin{figure}
\centering
\includegraphics[width=0.48\textwidth,bb=0 0 495 381]{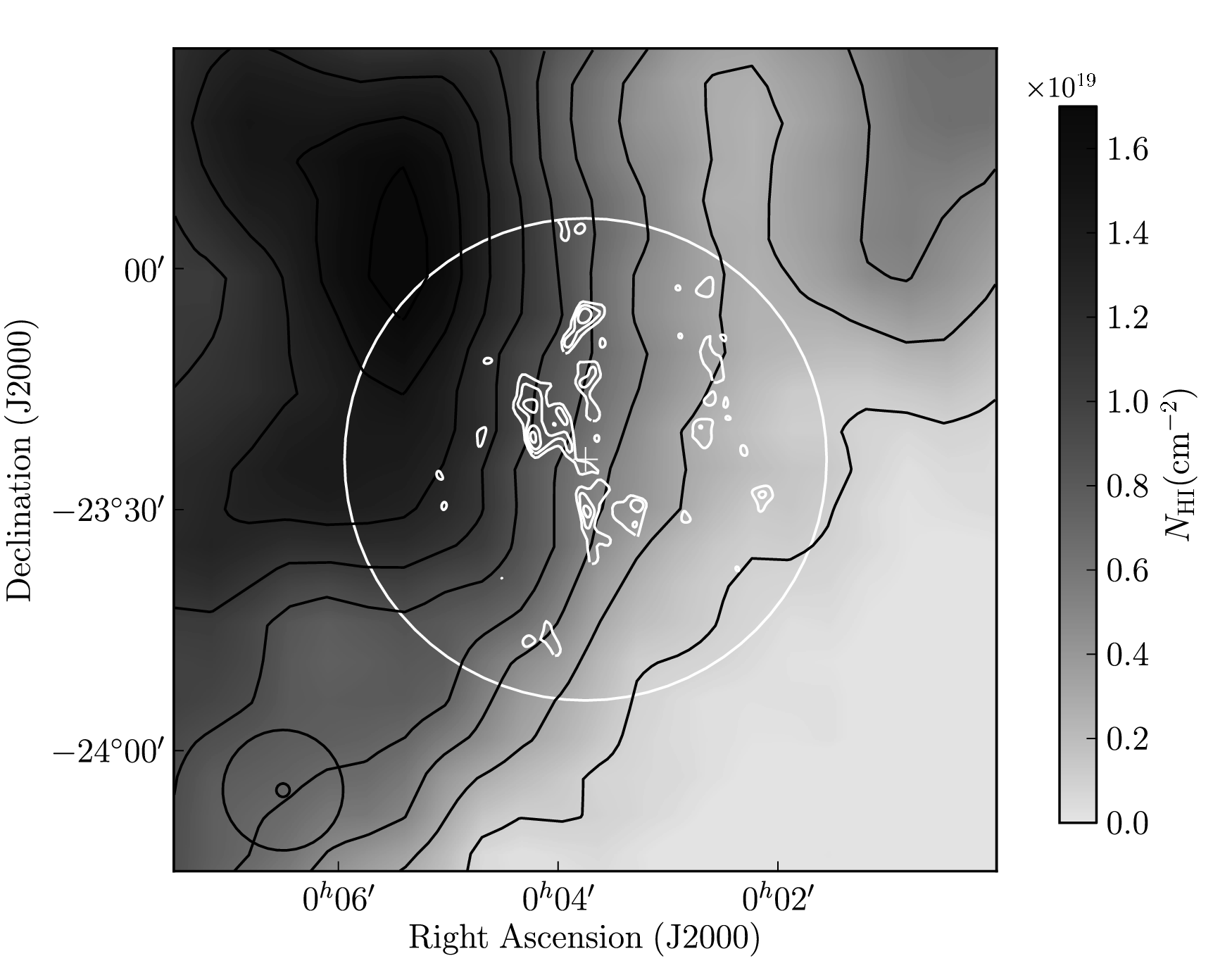}
\caption{\ion{H}{i} column density map in the direction of QSO\,J0003$+$2323 as observed with GASS, the black contours start at $N_\ion{H}{i} = 1\cdot 10^{18}\,\mathrm{cm}^{-2}$ and increase in steps of $2\cdot 10^{18}\,\mathrm{cm}^{-2}$.  Overlaid (white) are the contour lines of the high-resolution observation made with the VLA, starting at $N_\ion{H}{i} = 1\cdot 10^{18}\,\mathrm{cm}^{-2}$ in steps of $1\cdot 10^{18}\,\mathrm{cm}^{-2}$. In the lower left corner the beam sizes of EBHIS and the VLA (synthesized beam) are indicated. The line of sight towards QSO\,J0003$+$2323 is marked with a white cross, the white circle denotes the size of the primary beam of the VLA ($2\times\mathrm{HPBW}$).}
\label{ebhis_VLA}
\end{figure}

About 35\% of the \ion{Ca}{ii} and 20\% of the \ion{Na}{i} absorbers show multiple velocity components (e.g., Fig.\,\ref{spectrum}), suggesting the presence of substructure. Richter, Westmeier \& Br\"{u}ns (2005) and Ben Bekhti et al. (2009) observed five sight lines with the Very Large Array (VLA) and the Westerbork Synthesis Radio Telescope (WSRT) to search for such small-scale structures. In all five directions cold ($70< T_\mathrm{kin} <3700$\,K, corresponding to a linewidth of $\Delta v=1.8 \ldots 13$ km\,s$^{-1}$) and compact (down to sub pc scales) clumps were found embedded in a more diffuse environment; see the white contour maps in Fig.\,\ref{ebhis_WSRT} and \ref{ebhis_VLA} for two examples in the direction of QSO\,B1331$+$170 and QSO\,J0003$+$2323. 

For many of the sight lines we obtained maps of the \ion{H}{i} gas around the absorption systems using the data of EBHIS and GASS, showing both, diffuse and clumpy/filamentary extended structures. In addition to the interferometry data in Fig.\,\ref{ebhis_WSRT} and \ref{ebhis_VLA} we present the \ion{H}{i} column density maps observed with EBHIS and GASS. 

QSO\,B1331$+$170 shows a two-component structure in \ion{Ca}{ii} and \ion{Na}{i} absorption and \ion{H}{i} emission. The lines spread between $v_\mathrm{LSR}=-40 \ldots 0$\,km\,s$^{-1}$. 

In case of QSO\,J0003$+$2323 three \ion{Ca}{ii} absorption components are spread between $v_\mathrm{LSR}= -120 \ldots -80$\,km\,s$^{-1}$ and one broad 21-cm emission line is found within GASS at $v_\mathrm{LSR}= -112$\,km\,s$^{-1}$.

Surprisingly, there is no tight spatial correlation between the single-dish and interferometry data. While some structures seen in the high-resolution data are well traced by the EBHIS/GASS (e.g., the southern clump in Fig.\,\ref{ebhis_WSRT} bottom panel) other small-scale features are not visibly related to structures in the single-dish maps (e.g., north-western features in Fig.\,\ref{ebhis_WSRT} bottom panel). We will study this phenomenon in more detail in a future project (Ben Bekhti et al., in prep.).


\section{Conclusions and outlook}

The results presented show how important measurements at different wavelengths and with different resolutions are to get a more complete view of the properties of neutral halo gas. Such observations allow us to study a variety of elements, distinct gas phases, and structure sizes. \ion{H}{i} traces the galactic halo  with structures on all scales, from tens of AU to kiloparsecs in form of streams, clouds, tiny clumps, and filaments. 

In addition to the IVC/HVC clouds and extended complexes, the Milky Way halo contains a large number of low-column density absorbers with with typical \ion{H}{i} column densities of $1 \cdot 10^{18}\mathrm{cm}^{-2} \ldots 1 \cdot 10^{20}\,\mathrm{cm}^{-2}$. Cold and compact clumps (observed with radio synthesis telescopes) are embedded in a more diffuse envelope (observed with single-dish radio telescopes), which fits perfectly into the picture of a multiphase character of the halo gas. 

Assuming that the Milky Way  environment is typical for low redshift galaxies, weak \ion{Ca}{ii} absorption should arise in the neutral disks of galaxies and in their extended neutral gas halo. Richter, et al. (2010) searched for \ion{Ca}{ii} absorbers at low redshifts ($z < 0.5$). They detect 23 intervening \ion{Ca}{ii} absorbers out of 304 QSO sightlines with similar physical properties as around the Milky Way. In agreement with \ion{H}{i} observations they found that the radial extend of these halo absorbers around their host galaxies is about 55\,kpc. 

The similar CDD slopes of $\beta=-1.6$ for the \ion{Mg}{ii} (which traces the halos of other galaxies) and our \ion{Ca}{ii} absorbers indicate that the statistical properties of halo gas are comparable at low and high redshifts. 
All these observations lead to the conclusion that around disk galaxies there is a population of gaseous structures with very similar physical properties that likely influence the evolution of the host galaxy substantially.  

Although the understanding of halos made great progress in the recent years, there are many open questions. We know that most of the neutral gas seems to reside within 50\,kpc of the disks, but the true space distribution is largely unknown. Even for the nearby galaxies it is difficult to measure the neutral hydrogen down to low column densities. While absorption spectroscopy is much more sensitive it has the drawback that it needs suitable background sources, such that only a very incomplete picture of individual galaxies can be obtained. The determination of the mass distribution in the halos, however, is a key to quantify the gas accretion rate of galaxies and connect that to the observed constant star formation rate in disk galaxies of the Local Volume. The Square Kilometer Array (SKA, e.g., Garrett et al., 2010) and its pathfinders will hopefully allow us to answer this question.

Another issue is the observed lagging halo gas. Is it the result of internal (e.g., galactic fountains, winds) or external (accretion from the IGM) drivers? How do effects like star formation, accretion, and interaction between galaxies contribute to the total mass of extraplanar gas around galaxies? Are warps and lopsidedness observed in many disk galaxies directly related to the material in the halo? 

Little is known today when it comes to the smallest scales even for the large HVC complexes. Much of our knowledge today is based on survey data with poor angular resolution like the LAB. Recent studies show, that even with the new single-dish \ion{H}{i} surveys much more substructure is found revealing interesting new aspects. One example is the HVC complex GCN where EBHIS/GASS data resolve the previously known "large" clouds into tiny objects mostly even unresolved within the new survey data while no extended diffuse emission is detected (Winkel et al., 2010b; Winkel et al., in prep.). Will the observed fragmentation continue when going to even higher resolution and if so, what are the properties of this scaling?

The best way to confront all these open questions is to combine sensitive, high-resolution multi-wavelength data. Future instruments like SKA, the Atacama Large Millimeter Array (ALMA; e.g., Combes 2010), the Cosmic Origins Spectrograph (COS; Goudfrooij et al., 2010), the Extended Roentgen Survey with an Imaging Telescope Array (eRosita; Predehl, et al. 2010) will shed light on many of the open problems. 

From the theoretical point of view, numerical simulations can help to determine the complex dynamics of the multiphase ISM in the halos of galaxies. Especially, magneto-hydro\-dynamic simulations are the key to understand the interaction between the various gas phases in the ISM and magnetic fields. This will clarify whether the halo structures are stable objects supported by (large-scale) magnetic fields or if they are just transient objects in the turbulently mixed gas phases in the ISM.

\acknowledgements

The authors thank the Deutsche Forschungsgemeinschaft (DFG) for financial support under the research grant KE757/7-1 and KE757/9-1. We thank M. T. Murphy for providing the reduced VLT/UVES data which are the basis of our study. We thank P. M. W Kalberla, T. Westmeier, and Gyula J\'ozsa for their support and the fruitful scientific discussions.


\appendix

\end{document}